\documentclass[10pt,twoside]{mapas}

\usepackage[english]{babel}
\usepackage{amsmath,amsthm}
\usepackage{amsfonts}
\usepackage{graphicx}
\usepackage{tikz}
\usepackage{hyperref}

\theoremstyle{definition}

\theoremstyle{remark}

\usepackage[utf8]{inputenc}

\title{On the Non-Existence of Unbounded Discrete Space-Time}

\author[1]{\textbf{Furkan Semih Dündar} \thanks{the corresponding author: furkan.dundar@boun.edu.tr}}

\affil[1]{Boğaziçi University, Faculty of Arts and Sciences, Physics Department, İstanbul-Turkey} 

\shortauthor{Furkan Semih Dündar}

\author[1]{\textbf{Metin Arık}\thanks{metin.arik@boun.edu.tr}}

\shortauthor{Metin Arık}

\enabstract{$n$-scales are a generalization of time-scales that has been put forward to unify continuous and discrete analyses in higher dimensions. In this paper we investigate massive scalar field theory on a regular $n$-Scale. We have given the modified field equation that is appropriate to this space-time structure and gave the mode solutions. If the space-time is discrete we have found that no massive scalar field can exist. Hence, we concluded that unbounded discrete space-time cannot exist. If discrete space-time exists, it has to be bounded in each dimension.}
\enkeywords{
Time-Scale\\
$n$-Scale\\
Field Theory\\
Discrete Space-Time
}

\begin{document}
\maketitle

\section{Introduction}

In the Planck scale, it is believed that the spacetime has a granular structure. In order to explain the Planck scale physics, theoreticians put forward various theories. For example the causal dynamical triangulation approach triangulates the spacetime with filled-in cells \cite{cdt}, in the loop quantum gravity approach \cite{lqg} spacetime itself is discrete.

Apart from the discussions of quantum gravity, mathematicians have been working on the concept of time-scale. A time-scale is an arbitrary closed subset of $\mathbb R$ in the usual topology. For example the sets $[0,1]$, $\mathbb Z$, $\mathbb N$ or $[0,1]\cup\mathbb Z$ are all time scales. Time scale was developed in \cite{1,2,3}. Time-scale calculus unifies the discrete and continuous analyses. For a general overview one may see \cite{bulentogur}. However a time-scale is one dimensional and its multi-dimensional counterparts are in the form of product spaces \cite{bohner2005multiple}. This inadequacy in covering the real world applications, which may require non-product spaces, the concept of $n$-cale has been developed \cite{furkannScale}. The definition of an $n$-scale resembles that of a time scale: an $n$-Scale is an arbitrary closed subset of $\mathbb R^n$.

\section{Massive Scalar Field Theory on a Regular $n$-Scale}
\label{sec:gentheory}

\begin{figure}
    \centering
    \includegraphics[width=8cm]{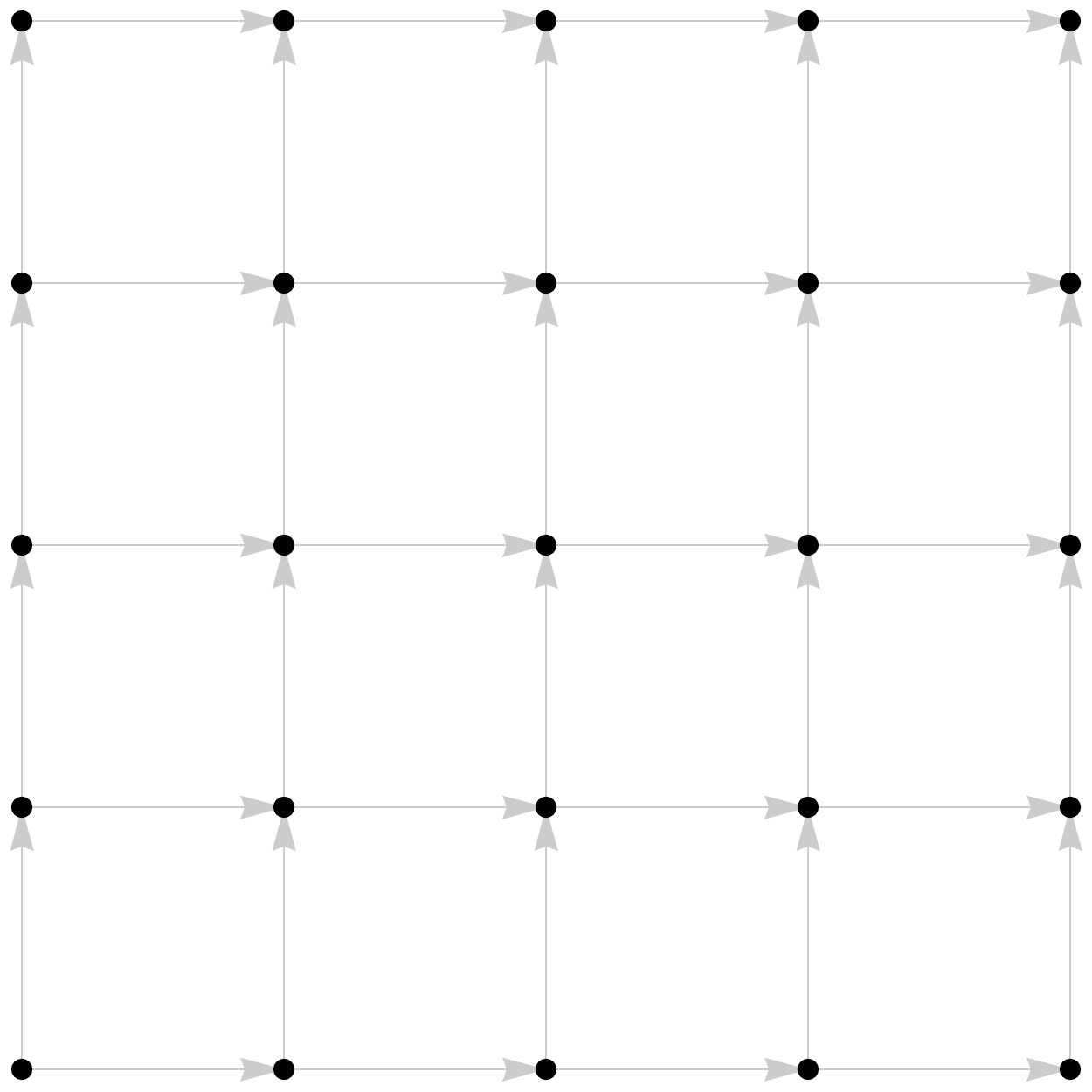}
    \caption{Part of a 2-scale in the form of $\alpha^0 \mathbb Z \otimes \alpha^1 \mathbb Z$ is depicted. Lattice spacing in the horizontal direction is $\alpha^1$, whereas it is $\alpha^0$ in the vertical direction. Arrows show the neighborhood structure of the 2-scale.}
    \label{fig:2-scale}
\end{figure}

In this section we give the Lagrangian density and derive the Euler-Lagrange equation using the $n$-Scale calculus. We give the Lagrangian density for a massive scalar field as follows:

\begin{equation}
    \mathcal L = \frac 12 \eta^{\mu\nu} \Delta_\mu \phi^{\rho^\mu} \Delta_\nu \phi^{\rho^\nu} - \frac 12 m^2 \phi^2,
\end{equation}

where $\Delta_\mu = \partial/\Delta x^\mu$ is the partial $\Delta$-derivative with respect to $x^\mu$ and the inverse metric is $\eta^{\mu\nu}=(+,-,-,\cdots,-)$. Because the $n$-Scale is regular (\emph{i.e.} in the form $\alpha^0 \mathbb Z \times \alpha^1 \mathbb Z \times \cdots \times \alpha^{n-1} \mathbb Z$, see Figure~\ref{fig:2-scale}), the forward jump operator ($\sigma^\mu$) is given by the following expression:

\begin{equation}
	\sigma^\mu(x) = x^\mu + \alpha^\mu,
\end{equation}

whereas the backward jump operator is given as:

\begin{equation}
	\rho^\mu(x) = x^\mu - \alpha^\mu.
\end{equation}

The action is then given by the following $\Delta$-integral:

\begin{equation}
    S = \int \prod_{\mu=0}^{n-1} \Delta x^\mu \mathcal L.
\end{equation}

For a definition of the integral see \cite{furkannScale}. The Euler-Lagrange equation is obtained by extremizing the action with respect to variations of the field and its $\Delta$-derivatives. Let us do the calculations:

\begin{align}
	\delta S &= \delta \int \prod_{j=0}^{n-1} \Delta x^j \left(\frac 12 \eta^{\mu\nu} \Delta_\mu \phi^{\rho^\mu} \Delta_\nu \phi^{\rho^\nu} - \frac 12 m^2 \phi^2\right),\\
    &= \int \prod_{j=0}^{n-1} \Delta x^j \left(\eta^{\mu\nu}\Delta_\mu \phi^{\rho^\mu} \Delta_\nu \delta \phi^{\rho^\nu} - m^2\phi \delta\phi\right),\\
    \intertext{Then by using the integration by parts technique \cite{var-calc} (supposing that $\delta \phi$ vanishes at infinity or on the boundary), we obtain:}
    &= -\int \prod_{j=0}^{n-1} \Delta x^j \left(\eta^{\mu\nu} \Delta_\mu \Delta_\nu \phi^{\rho^\nu} + m^2\phi\right)\delta\phi,\\
    &= 0.
\end{align}

Therefore, the Euler-Lagrange equation we obtain is the following:

\begin{equation}
	\eta^{\mu\nu} \Delta_\mu \Delta_\nu \phi^{\rho^\nu} + m^2\phi = 0.\label{eq:eommsf}
\end{equation}

The solution of this equation is given by a product of exponential functions on each time scale that is a part of the regular $n$-scale. In \cite{bulentogur} there is listed a useful identity of time scale exponential function:

\begin{equation}
	e_p(\sigma(t),s) = (1+\mu(t)p(t))e_p(t,s).\label{eq:epsigma}
\end{equation}

Let $p$ be a constant ($p$ is usually a function that appears in the equation that defines the time scale exponential function through $\Delta_t e_p(t,s) = p(t) e_p(t,s)$) and the graininess function $\mu$ is defined through $\mu(t) = \sigma(t)-t$. In our case we have $\mu^\mu(x) = \sigma^\mu(x)-x^\mu = \alpha^\mu$. Therefore by generalizing Equation~(\ref{eq:epsigma}) to higher dimensions and applying $\rho$ (backward jump operator) to variable $t$, we obtain:

\begin{align}
	e_p(t,s) &= \frac{1}{1+\mu(t)p(t)}e_p(\sigma(t),s)\\
    e_p(\rho(t),s) &= \frac{1}{1+\mu(\rho(t))p(\rho(t))}e_p(\sigma(\rho(t)),s),\\
    e_{p_j}(\rho^j(x^j),s^j) &= \frac{1}{1+\alpha^j p_j}e_{p_j}(x^j,s^j),
\end{align}

where there is no sum over $j$. In Cartesian coordinates, the mode solutions of the field equation~(\ref{eq:eommsf}) are given by the exponential function on $n$-scales (we let $s^j = $ for convenience):

\begin{equation}
    \phi_{\vec{k}} = \prod_{j=0}^{n-1} e_{ik_j}(x^j,0),\label{eq:phigensol}
\end{equation}

with the condition that 

\begin{equation}
	\frac{k_0^2}{1+i\alpha^0k_0}-\frac{k_1^2}{1+i\alpha^1k_1}-\cdots -\frac{k_{n-1}^2}{1+i\alpha^{n-1}k_{n-1}} = m^2.\label{eq:keqn}
\end{equation}

For definition of the function $e_{\cdot}(\cdot)$ see \cite{bulentogur}. This type of solution is valid for an $n$-scale in the product form of $\mathbb T_1 \otimes \mathbb T_2 \otimes \cdots \mathbb T_n$ where each $\mathbb T_i$ are unbounded 1-scales. If one or many of the 1-scales are bounded, either from below or above or both, then boundary conditions should be imposed to find the mode solutions as superposition of Equation~(\ref{eq:phigensol}).

In the continuous case where the $n$-Scale is $\mathbb R^n$, we have $\forall \mu, \alpha^\mu = 0$. Moreover the exponential function becomes the usual one $e^{ik_jx^j}$ (no sum over $j$). Hence the equation relating the wave-numbers and the mass is the usual $k_0^2 - k_1^2 -\cdots k_{n-1}^2 = m^2$. Our model produces the correct mode solution in the continuum case.

In the discrete case where $\forall \mu, \alpha^\mu \neq 0$ the solutions of Equation~(\ref{eq:keqn}) are in general complex. Let $ik_\mu = \kappa_\mu + i\kappa_\mu^*$. Then the mode solution consists of two parts: 1) wave-like, 2) exponentially diverging:

\begin{equation}
	\phi_{\vec{k}} = \prod_{\mu=0}^{n-1} e_{i\kappa_\mu^*}(x^\mu,0) \prod_{\nu=0}^{n-1} e_{\kappa_\nu}(x^\nu,0).
\end{equation}

For constant graininess function $\mu(t)=\alpha$ and $p$, we have:

\begin{equation}
	e_p(t,0) = (1+\alpha p)^{t/\alpha}.
\end{equation}

Therefore when the parameter $p$ of the exponential function is pure imaginary we have oscillating solutions and when it is real we have exponentially diverging solutions. What is more, even in the oscillating case the modulus of the mode solution diverges exponentially. Hence if a solution of massive scalar field theory should be regarded as physical, all of the $k_\mu$'s should be zero. This is the only sensible solution but it does not have dynamics. Therefore we conclude that if the space-time is discrete and unbounded there is no sensible physical solution. Henceforth we arrive an important conclusion that if space-time is an $n$-Scale in the form of ($\alpha^0 \mathbb Z \times \alpha^1 \mathbb Z \times \cdots \times \alpha^{n-1} \mathbb Z$) it cannot exist physically. (Because in nature there is a massive scalar field whose particle is famously called the Higgs boson, although it is not a free field as considered in this paper.) Notice that this implies time/space is either a continuous variable or is bounded from below and above. We hope that our result for a specific case of an $n$-Scale can be generalized to more generic discrete space-times.

\section{Conclusion}
\label{sec:conc}

In this paper we investigated massive scalar field theory on a regular $n$-Scale which is the product of time-scales in the form of $\alpha \mathbb Z$ where $\alpha$ is a positive real number. We have given the field equations and their mode solutions using the exponential functions that are defined on each time-scale that constitute the parts of the general $n$-Scale.

In the case where the space-time is continuous our model predicted the correct form of the solutions that are well-known in the literature. However, when there is discreteness at least in one dimension (either space or time) our model gave exponentially diverging solutions. Hence we concluded that if the space-time is discrete and can be written as a product of time-scales in the form of $\alpha^0 \mathbb Z \times \alpha^1 \mathbb Z \times \cdots \times \alpha^{n-1} \mathbb Z$ it cannot exist physically. Because there is a massive scalar field in nature called as the Higgs boson although it is not a free field as considered in this paper, we rule out the existence of an unbounded discrete space-time. The only possibility is that the space-time is bounded by a boundary if it is discrete. Only continuous unbounded space-time is possible (\emph{e.g.} Minkowski space-time).

\section{Acknowledgements}

The authors are thankful to Bayram Sözbir for useful discussions and to an anonymous referee for constructive criticisms. F.S.D. is supported by TUBİTAK 2211 Scholarship. The research of F.S.D. and M.A. is partly supported by the research grant from Boğaziçi University Scientific Research Fund (BAP), research project No. 11643.

\bibliographystyle{plain}
\bibliography{refs}

\end{document}